\documentclass[aps,twocolumn,prl,showpacs]{revtex4}
\usepackage[dvips]{graphicx}
\usepackage{graphics}
\usepackage{dcolumn}
\usepackage{amssymb}
\usepackage{bm}
\usepackage{amsmath}

\def\spose#1{\hbox to 0pt{#1\hss}}

\def\lta{\mathrel{\spose{\lower 3pt\hbox{$\mathchar"218$}}
     \raise 2.0pt\hbox{$\mathchar"13C$}}}
\def\gta{\mathrel{\spose{\lower 3pt\hbox{$\mathchar"218$}}
     \raise 2.0pt\hbox{$\mathchar"13E$}}}
\newcommand{\be}{\begin{equation}}
\newcommand{\en}{\end{equation}}
\newcommand{\bea}{\begin{eqnarray}}
\newcommand{\ena}{\end{eqnarray}}

\newcommand{\dd}{\mbox{d}}

\def\setR{\mathbb{R}}

\def\ii{\textrm i}
\def\ee{\textrm e}

\newcommand{\dbb}{de$\,$Broglie-Bohm}

\begin{document}
\title{Quantum-to-classical transition of primordial cosmological perturbations in de Broglie--Bohm quantum theory}

\author{Nelson Pinto-Neto}
\affiliation{ICRA - Centro Brasileiro de
Pesquisas F\'{\i}sicas -- CBPF, rua Xavier Sigaud, 150, Urca,
CEP22290-180, Rio de Janeiro, Brazil}

\author{Grasiele Santos}
\affiliation{ICRA - Centro Brasileiro de
Pesquisas F\'{\i}sicas -- CBPF, rua Xavier Sigaud, 150, Urca,
CEP22290-180, Rio de Janeiro, Brazil}

\author{Ward Struyve}
\affiliation{Institute for Theoretical Physics, K.U.Leuven,
Celestijnenlaan 200D, B--3001 Leuven, Belgium}
\affiliation{Institute of Philosophy, K.U.Leuven, Kardinaal Mercierplein 2, B--3000 Leuven, Belgium.}

\date{\today}

\begin{abstract}

There is a widespread belief that the classical small inhomogeneities which gave rise to all structures in the Universe through gravitational instability originated from primordial quantum cosmological fluctuations. However, this transition from quantum to classical fluctuations is plagued with important conceptual issues, most of them related to the application of standard quantum theory to the Universe as a whole. In this paper, we show how these issues can easily be overcome in the framework of the de$\,$Broglie-Bohm quantum theory. This theory is an alternative to standard quantum theory that provides an objective description of physical reality, where rather ambiguous notions of measurement or observer play no fundamental role, and which can hence be applied to the Universe as a whole. In addition, it allows for a simple and unambiguous characterization of the classical limit.

\end{abstract}


\maketitle

Presently, there is a vivid discussion in the literature about conceptual issues concerning the transition from primordial cosmological quantum fluctuations to the small classical inhomogeneities which gave rise to the structures in the Universe, such as stars and galaxies~\cite{guth85,albrecht94,polarski96,lesgourgues97,kiefer98,kiefer09,burgess08,perez06,sudarsky11}.

All cosmological pictures of structure formation, either inflationary~\cite{liddle00,mukhanov05,weinberg08,lyth09,peter09} or bouncing models~\cite{novello08}, rely on the fact that in its far past the Universe became so homogeneous and isotropic that only quantum vacuum fluctuations of inhomogeneities could have survived. Initially, the modes of the fluctuations of cosmological interest have physical wavelengths much smaller than the curvature scale of the background, and their quantum state is very close to the Minkowski vacuum. In the course of cosmological evolution, the physical wavelengths of these fluctuations become of the size of the curvature scale, at which point they begin to feel the background gravitational field. Once the wavelengths are much bigger than the curvature scale, the fluctuations are believed to become classical. They then give rise to the structures in the Universe, through gravitational instability.

The quantum-to-classical transition of the fluctuations is often seen as a result of the squeezing of the vacuum state. It is argued that the squeezed state is somehow equivalent to an element of an ordinary ensemble of classical fields (see e.g.\ \cite{guth85,albrecht94,polarski96,lesgourgues97,kiefer98,kiefer09,burgess08,weinberg08}). The argument often invokes decoherence, which is assumed to lead to loss of possible quantum interference for degrees of freedom of interest, due to their interaction with other degrees of freedom.

However, as pointed out by Sudarsky and collaborators~\cite{perez06,sudarsky11}, the usual accounts of the quantum-to-classical transition have serious shortcomings. They argued that no satisfactory justification has been given on why the quantum state, which is translational and rotational invariant, and which remains so during Schr\"odinger time evolution, results in a noninvariant state. Even when there is suitable decoherence, which suppresses interference between different noninvariant terms into which the quantum state can be decomposed, it is not explained why one of these terms is selected. According to standard quantum theory, a transition to a noninvariant state could only be obtained by collapse, which is supposed to happen upon a measurement by an external observer or measurement device. For instance, in the case of a decaying atom emitting a photon with a spherically symmetric wave function, one may consider an external measurement device which will bring about the collapse of the photon's wave function so that it will be detected in a particular direction. However, in the cosmological context, there is something qualitatively different: we are dealing with cosmological primordial fluctuations which will give rise to all the structures in the Universe. There can be no measurement device external to it. All the structures, including measurement devices, are supposed to emerge from the primordial fluctuations themselves. As such, in the absence of an external device, it is unclear how the translational and rotational invariance of the quantum state of the fluctuations can be broken. This is the problem with the cosmological scenario, already acknowledged in~\cite[p.\ 64]{liddle00}, \cite[p.\ 348]{mukhanov05} and \cite[pp.\ 386-387]{lyth09}.

Sudarsky and collaborators~\cite{perez06,deunanue08,leon10,sudarsky11} proposed to solve this problem in the context of spontaneous collapse theories, in which the collapse is an unambiguous, objective stochastic process. The whole framework is still under construction and may yield possible deviations from the standard predictions.

In this paper, we will address the problem within the \dbb\ theory~\cite{bohm93,holland93b,durr09}. In this theory, the Universe is described by the universal wave function, together with an actual configuration for gravity (e.g.\ an actual three-metric) and a configuration for the matter (e.g.\ particle positions or fields). The universal wave function should satisfy the appropriate quantum cosmological wave equation. Its role is to determine the dynamics of the actual configuration, which is deterministic.

For the problem at hand, we will not consider the full quantum cosmological framework, but instead restrict ourselves to the effective theory for the perturbations in terms of the Mukhanov-Sasaki variable, which represents a gauge invariant combination of perturbations of the metric and inflaton field {\footnote{In the \dbb\ approach, it is possible to obtain the effective theory of perturbations from the full quantization of the background geometry and linear perturbations, see~\cite{falciano09}.}}. Although the quantum state remains translational and rotational invariant in this framework, the actual \dbb\ field corresponding to the perturbations, which is not present in the standard approach, breaks this symmetry. The initial field configuration is not determined by the theory, but in quantum equilibrium it is distributed according to the quantum mechanical distribution.

Also the question of the classical limit is straightforward in the \dbb\ theory: the classical limit is obtained whenever the actual field configuration evolves approximately according to the classical equations. We will see that the actual \dbb\ perturbations reach the classical limit at the expected stage and that the quantum equilibrium ensemble of these perturbations then corresponds exactly to the classical ensemble that is assumed in~\cite{guth85,albrecht94,polarski96,lesgourgues97,kiefer98,kiefer09}.

We will focus on the usual inflationary models with one scalar field $\varphi$ moving under a potential. The background spatial metric is assumed to be flat. Considering only scalar perturbations (tensor perturbations could be considered in a similar way), the metric up to first order can be written in the longitudinal gauge as
\be
\dd s^2 = a^2(\eta) \left\{ \left[1+2\phi(\eta,{\bf x})\right]\dd\eta^2 -
\left[1-2\phi(\eta,{\bf x})\right] \delta_{ij} \dd x^i \dd x^j \right\},
\en
where $\phi$ is the Bardeen potential in this gauge, and $\eta$ is the conformal time. There is also a perturbation of the inflaton field $\delta \varphi (\eta,{\bf x})$. Because of the general relativity constraint equations, there is only one degree of freedom left, which can be described by the gauge invariant Mukhanov-Sasaki variable $y(\eta,{\bf x})$ given by
\be
y  \equiv a\left[\delta\varphi + \frac{\varphi '}{\cal{H}}\phi\right],
\en
where ${\cal{H}}=a'/a$, with $a$ the scale factor. A prime denotes a derivative with respect to $\eta$.

It is convenient to introduce Fourier modes, through
\begin{equation}
\label{mode}
y(\eta, {\bf x})=\int{\frac{d^3k}{(2\pi)^{3/2}}y_{\bf k}(\eta) \ee^{\ii {\bf k} \cdot {\bf x}}},
\end{equation}
where $y^*_{\bf k} = y_{-{\bf k}}$ (due to the reality of $y(\eta, {\bf x})$). The Hamiltonian for these modes, as implied by general relativity, reads
\begin{equation}
\label{hk}
H = \int_{\setR^{3+}} d^{3}k \left[p_{\bf k}p_{\bf k}^* + k^2 y_{\bf k}y_{\bf k}^* +
\frac{z'}{z}(p_{\bf k}y_{\bf k}^* + y_{\bf k}p_{\bf k}^*)\right],
\end{equation}
where $z\equiv 2\sqrt{\pi}a\varphi '/(m_{Pl}{\cal{H}})$, $k = |{\bf k}|$, and where $p^*_{\bf k}$ is the momentum conjugate to $y_{\bf k}$. Note that the integration is restricted to half the number of possible modes (denoted by $\setR^{3+}$), so that only independent ones are included. The corresponding equations of motion are
\be
\label{modeeq}
y_{\bf k}'' +  \left(k^2-\frac{z''}{z}\right) y_{\bf k}=0.
\en

Let us first analyze this classical theory. At the onset of inflation, physical modes are assumed to have wavelengths much smaller than the curvature scale, i.e., $k^2 \gg z''/z$, so that their equation of motion approximately reduces to that of a free mode in Minkowski space-time. In many inflationary models (like power-law inflation or slow-roll~\cite{peter09}), the precise behavior of these physical modes at early times ($\eta \to \eta_i$, where $\eta_i$ is some initial time, with $|\eta_i| \gg 1$), is given by
\begin{equation}
\label{fka}
y_{\bf k}(\eta) \sim \ee^{- \ii k \eta}\left(1 + \frac{A_k}{\eta} + \dots \right).
\end{equation}
The physical modes will grow larger during inflation and will eventually obtain wavelengths much bigger than the curvature scale, i.e., $k^2 \ll z''/z$. At that stage, in many models, the modes are approximately given by
\begin{equation}
\label{yqq}
y_{\bf k} (\eta) \sim A^d_k \eta^{\alpha_d} + A^g_k \eta ^{\alpha_g} \approx A^g_k \eta ^{\alpha_g},
\end{equation}
where $\alpha_d >0$ and $\alpha_g <0$. The first term represents the decaying mode and the second one the growing mode, which dominates in $y_{\bf k}$.

Let us now turn to the corresponding quantum theory. The details of the quantization can for example be found in~\cite{polarski96}. In the functional Schr\"odinger picture, the state is a wave functional $\Psi(y,y^*, \eta)$ and satisfies a functional Schr\"odinger equation determined by the quantum Hamiltonian corresponding to~\eqref{hk}. In the special case the wave functional is a product $\Psi = \Pi_{{\bf k} \in \setR^{3+}} \Psi_{\bf k}(y_{\bf k},y^*_{\bf k},\eta)$, like for the vacuum, each mode wave function satisfies the Schr\"odinger equation (note that we formally treat $y_{\bf k},y^*_{\bf k}$ as independent fields; equivalently, their real and imaginary components can
be used, defined by $y_{\bf k} = (y_{\bf k,r} + \ii y_{\bf k,i})/{\sqrt 2}$):
\begin{widetext}
\begin{equation}
\label{sch}
\ii\frac{\partial\Psi_{\bf k}}{\partial\eta}=
\left[ -\frac{\partial^2}{\partial y_{\bf k}^*\partial y_{\bf k}}+
k^2 y_{\bf k}^* y_{\bf k}
- \ii\frac{z'}{z}\left(\frac{\partial}{\partial y_{\bf k}^*}y_{\bf k}^*+
y_{\bf k}\frac{\partial}{\partial y_{\bf k}}\right)\right]\Psi_{\bf k}.
\end{equation}
For the ground state, the mode wave functions are given by (see Ref.~\cite{polarski96})
\begin{equation}
\label{psi2}
\Psi_{\bf k} = \frac{1}
{\sqrt{\sqrt{2\pi}|f_k(\eta)|}} \exp{\left\{-\frac{1}{2|f_k(\eta)|^2}|y_{\bf k}|^2 +   \ii \left[\left(\frac{|f_k(\eta)|'}{|f_k(\eta)|}-
\frac{z'}{z}\right)|y_{\bf k}|^2-
\int^\eta \frac{d {\tilde \eta}}{2|f_k({\tilde \eta})|^2}\right]\right\}} ,
\end{equation}
with $f_k$ a solution to the classical mode equation~\eqref{modeeq}, and $f_k(\eta_i) = 1/\sqrt{2k}$, where $|\eta_i|\gg1$. This state is homogeneous and isotropic.

In order to pass to the \dbb\ approach, we write $\Psi_{\bf k} = R_{\bf k} \ee^{\ii S_{\bf k}}$, with $R_{\bf k}=|\Psi_{\bf k}|$, so that the Schr\"odinger equation~\eqref{sch} yields the two real equations
\begin{equation}
\label{h-j}
\frac{\partial S_{\bf k}}{\partial \eta} + \frac{\partial S_{\bf k}}{\partial y^*_{\bf k}}\frac{\partial S_{\bf k}}{\partial y_{\bf k}}+
k^2y^*_{\bf k}y_{\bf k}+\frac{z'}{z}\left(\frac{\partial S_{\bf k}}{\partial y^*_{\bf k}}y^*_{\bf k}+
y_{\bf k}\frac{\partial S_{\bf k}}{\partial y_{\bf k}}\right)-\frac{1}{R_{\bf k}}\frac{\partial^2R_{\bf k}}{\partial y^*_{\bf k}\partial y_{\bf k}} =0 ,
\end{equation}
\begin{equation}
\label{cont}
\frac{\partial R_{\bf k}^2}{\partial \eta}+
\frac{\partial}{\partial y_{\bf k}}\left[R_{\bf k}^2\left(\frac{\partial S_{\bf k}}{\partial y^*_{\bf k}}+\frac{z'}{z}y_{\bf k}\right)\right]+
\frac{\partial}{\partial y^*_{\bf k}}\left[R_{\bf k}^2\left(\frac{\partial S_{\bf k}}{\partial y_{\bf k}}+\frac{z'}{z}y^*_{\bf k}\right)\right] =0 .
\end{equation}
\end{widetext}
We can then postulate an actual field $y(\eta,{\bf x})$, whose modes obey the \dbb\ {\em guidance equations}
\begin{equation}
\label{guidance}
y'_{\bf k}= \frac{\partial S_{\bf k}}{\partial y^*_{\bf k}}+\frac{z'}{z}y_{\bf k},  \quad
{y^*_{\bf k}}'= \frac{\partial S_{\bf k}}{\partial y_{\bf k}}+\frac{z'}{z}y^*_{\bf k} .
\end{equation}
By Eq.~\eqref{cont}, these equations of motion guarantee that the mode distribution $R^2_{\bf k}=|\Psi_{\bf k}|^2$ is preserved over time (and this is exactly the motivation to introduce these equations of motion). That is, if the initial distribution of the field modes is given by $|\Psi_{\bf k}|^2$, then it will be given by $|\Psi_{\bf k}|^2$ at any time. The particular distribution $R^2_{\bf k}=|\Psi_{\bf k}|^2$ plays the role of an equilibrium distribution and is called the {\em quantum equilibrium distribution}.

The guidance equations also follow from the Hamilton equation $y'_{\bf k} = p_{\bf k} +\frac{z'}{z}y_{\bf k}$ corresponding to the classical Hamiltonian \eqref{hk}, with $p_{\bf k}$ replaced by $\frac{\partial S_{\bf k}}{\partial y^*_{\bf k}}$. Equation \eqref{h-j} then resembles the Hamilton-Jacobi equation corresponding to the classical Hamiltonian, with an extra potential
\be
Q_{\bf k} \equiv - \frac{1}{R_{\bf k}}\frac{\partial^2R_{\bf k}}{\partial y^*_{\bf k}\partial y_{\bf k}},
\en
called the {\em quantum potential}.

Taking the time derivative of these equations and using~\eqref{h-j}, we obtain
\be
y_{\bf k}'' +  \left(k^2-\frac{z''}{z}\right) y_{\bf k} = - \frac{\partial Q_{\bf k}}{\partial y^*_{\bf k}}.
\en
This is just the classical equation of motion~\eqref{modeeq}, modified with an additional {\em quantum force}. The classical limit is obtained when this quantum force is negligible.

For the ground state, the guidance equations are easily integrated and yield
\be
\label{soly}
y_{\bf k}(\eta) =  y_{\bf k}(\eta_i)\frac{|f_k(\eta)|}{|f_k(\eta_i)|}.
\en
Note that this result is independent of the precise form of $f_k(\eta)$ and hence is quite general. In quantum equilibrium, the probability distribution of the initial configuration $y_{\bf k}(\eta_i)$ is given by $|\Psi_{\bf k} \left( y_{\bf k}(\eta_i),y^*_{\bf k}(\eta_i),\eta_i \right)|^2$.

For physical wave lengths and $\eta \rightarrow \eta_i$, the behavior of $f_k(\eta)$ is given by Eq.~(\ref{fka}). As such,
\be
y_{\bf k}(\eta) \sim \left( 1+\frac{{\rm Re} A_k}{\eta} +  \dots \right)
\en
(in many inflationary scenarios, ${\rm Re} A_k = 0$ and the first order term disappears). So $y_{\bf k}$ tends to be time independent for $|\eta|\gg1$. (This is compatible with the fact that the \dbb\ field configuration is stationary for the ground state of a quantized scalar field in Minkowski space-time~\cite{holland93b}.) Hence, the time dependence of the \dbb\ field configuration is completely different from that of classical solutions, which oscillates for $|\eta|\gg 1$ and $k^2\gg z''/z$, see Eq.~(\ref{fka}).

When the wavelengths become much bigger than the curvature scale ($k^2 \ll z''/z$), the behavior of $f_k(\eta)$ is approximately given by the growing mode, see Eq.~(\ref{yqq}), so that $|f_k|$ equals $f_k$, up to a time-independent complex factor. As such, the \dbb\ field configuration approximately evolves according to the classical field equation~\eqref{modeeq}, so that the classical limit has been attained.

The classical limit can also be investigated by examining the behavior of the quantum force and leads to the same result. For the ground state~(\ref{psi2}), the quantum force is given by
\begin{equation}
\label{Fpert1}
F_{Q,{\bf k}} \equiv - \frac{\partial Q_{\bf k}}{\partial y_{\bf k}^*} =  \frac{y_{\bf k}}{4|f_k|^4}
\end{equation}
for the mode ${\bf k}$. The classical force can be read from Eq.~(\ref{modeeq}) and the ratio is
\begin{equation}
\label{FCpert2}
\frac{F_{C,{\bf k}}}{F_{Q,{\bf k}}} =  - 4|f_k|^4\left(k^2 - \frac{z''}{z}\right).
\end{equation}
For $k^2 \gg \frac{z''}{z}$, this ratio is approximately $-1$ so that the quantum force cancels the classical force and the mode evolves freely. The guidance equations further restrict the velocities to be zero, so that the mode stands still. For $k^2 \ll \frac{z''}{z}$, this ratio becomes very big because of the growing mode, so that the quantum force becomes negligible with respect to the classical force. As a result, the mode will evolve according to the classical equation of motion. In this way, the transition from quantum to classical behavior is clear and simple.

Let us now turn to the statistical predictions. First, let us denote $y(\eta,{\bf x};y_i)$, with $y_i$ a field on space, a solution to the guidance equations such that $y(\eta_i,{\bf x};y_i) = y_i({\bf x})$. As noted before, if the initial field $y_i$ is distributed according to quantum equilibrium, i.e., $|\Psi(y_i,\eta_i)|^2$, then $y(\eta,{\bf x};y_i)$ will be distributed according to $|\Psi(y,\eta)|^2$. For such an equilibrium ensemble, we can consider the two-point correlation function
\begin{align}
&\left\langle y(\eta,{\bf x})y(\eta,{\bf x}+{\bf r})\right\rangle_{\rm dBB}  \\
& \quad = \int \mathcal{D} y_i |\Psi(y_i, \eta_i)|^2 y(\eta,{\bf x};y_i) y(\eta,{\bf x} + {\bf r};y_i) \\
& \quad = \int \mathcal{D} y |\Psi(y, \eta)|^2 y({\bf x}) y({\bf x} + {\bf r})
\end{align}
which is the usual expression for the correlation function, and can be calculated to yield
\be
\left\langle y(\eta,{\bf x})y(\eta,{\bf x}+{\bf r})\right\rangle_{\rm dBB} = \frac{1}{2\pi^2}\int dk  \frac{\sin{kr}}{r} k |f_{k}(\eta)|^2 ,
\en
in the case of the ground state. Just as in the usual account, this ensemble average should approximately be equal to the spatial average of an actual field configuration for the universe. This could be justified by adopting the usual assumption that the spatial integral can be taken over a larger volume than that over which the fields are correlated~\cite{liddle00,mukhanov05,weinberg08}.

To conclude, we have shown that the quantum-to-classical transition is obtained in the context of the \dbb\ theory. We have only considered scalar perturbations. Similar results apply to tensor perturbations. In this transition no appeal was made to decoherence. If at some stage in this transition, there is decoherence in the field basis, this will not destroy the classical behavior of the fields.

\begin{acknowledgments}
N.P.N.\ and G.S.\ would like to thank CNPq of Brazil and G.S.\ would like to thank Faperj for financial support. W.S.\ is funded by the Fund for Scientific Research--Flanders (FWO--Vlaanderen). W.S.\ is grateful to CBPF, where part of this work was carried out, for the hospitality. N.P.N.\ and W.S.\ are grateful to J\'er\^ome Martin and Antony Valentini, respectively, for introducing them to the problem of the quantum-to-classical transition in inflation theory. N.P.N.\ is also grateful to Daniel Sudarsky for some illuminating discussions on this problem.
\end{acknowledgments}

\end{document}